\RequirePackage{ifpdf}
\ifpdf % We are running pdfTeX in pdf mode
\documentclass[pdftex]{sigma}
\else
\documentclass{sigma}
\fi

\begin{document}

\allowdisplaybreaks

\renewcommand{\PaperNumber}{041}

\FirstPageHeading

\renewcommand{\thefootnote}{$\star$}

\ShortArticleName{Integrable String Models in Terms of Chiral
Invariants of $SU(n)$, $SO(n)$, $SP(n)$ Groups}

\ArticleName{Integrable String Models in Terms\\ of Chiral
Invariants of $\boldsymbol{SU(n)}$, $\boldsymbol{SO(n)}$,
$\boldsymbol{SP(n)}$ Groups\footnote{This paper is a contribution
to the Proceedings of the Seventh International Conference
``Symmetry in Nonlinear Mathematical Physics'' (June 24--30, 2007,
Kyiv, Ukraine). The full collection is available at
\href{http://www.emis.de/journals/SIGMA/symmetry2007.html}{http://www.emis.de/journals/SIGMA/symmetry2007.html}}}

\Author{Victor D. GERSHUN}

\AuthorNameForHeading{V.D. Gershun}

\Address{ITP, NSC Kharkiv Institute of Physics and Technology,
Kharkiv, Ukraine}
\Email{\href{mailto:gershun@kipt.kharkov.ua}{gershun@kipt.kharkov.ua}}

\ArticleDates{Received October 30, 2007, in f\/inal form April 22,
2008; Published online May 06, 2008}

\Abstract{We considered two types of string models: on the
Riemmann  space of string coordinates with null torsion and on the
Riemman--Cartan space of string coordinates with constant torsion.
 We used the hydrodynamic approach of Dubrovin, Novikov to
integrable systems and Dubrovin solutions of WDVV associativity
equation to construct new integrable string equations of
hydrodynamic type on the torsionless Riemmann space of chiral
currents in f\/irst case.
 We used the invariant local chiral currents of principal chiral
models for $SU(n)$, $SO(n)$, $SP(n)$ groups to construct new
integrable string equations of hydrodynamic type on the Riemmann
space of the chiral primitive invariant currents and on the chiral
non-primitive Casimir operators as Hamiltonians in second case.
 We also used Pohlmeyer tensor nonlocal currents to construct new
nonlocal string equation.}


\Classification{81T20; 81T30; 81T40; 37J35; 53Z05; 22E70}

\section{Introduction}

  String theory is a very promising candidate for a unif\/ied quantum theory of
gravity and all the other forces of nature. For quantum
description of string model we must have classical solutions of
the string in the background f\/ields. String theory in suitable
space-time backgrounds can be considered as principal chiral
model. The integrability of the classical principal chiral model
is manifested through an inf\/inite set of conserved charges,
which can form non-Abelian algebra. Any charge from the commuting
subset of charges and any Casimir operators of charge algebra can
be considered as Hamiltonian in bi-Hamiltonian approach to
integrable models. The bi-Hamiltonian approach to integrable
systems was initiated by Magri \cite{ger:Mag}.
  Two Poisson brackets~(PBs)
\begin{gather*}\{ \phi^{a}(x), \phi^{b}(y)\}_{0}=P^{ab}_{0}(x,y)
(\phi ),\qquad
\{\phi^{a}(x), \phi^{b}(y)\}_{1}=P^{ab}_{1}(x,y)(\phi)%\label{eq1}
\end{gather*}
are called compatible if any linear combination of these PBs
\[ \{ * , * \}_{0}+\lambda \{ * , * \}_{1}\]
is PB also for arbitrary constant $\lambda$. The functions
$\phi^{a}(t,x)$, $a=1,2,\dots,n$ are local coordinates on a
certain given smooth $n$-dimensional manifold $M^{n}$. The
Hamiltonian operators $ P^{ab}_{0}(x,y)(\phi)$,
$P^{ab}_{1}(x,y)(\phi)$ are functions of local coordinates
$\phi^{a}(x)$. It is possible to f\/ind such
Hamilto\-nians~$H_{0}$ and $H_{1}$ which satisfy bi-Hamiltonian
condition \cite{ger:Das}
\begin{gather*}%\label{eq2}
\frac{d\phi^{a}(x)}{dt}=\{\phi^{a}(x), H_{0}\}_{0}= \{
\phi^{a}(x),H_{1}\}_{1},\end{gather*} where
$H_{M}=\int_{0}^{2\pi}h_{M}(\phi(y))dy$, $M=0,1$. Two branches of
hierarchies arise under two equations of motion under two
dif\/ferent parameters of evolution $t_{0M}$ and $t_{M0}$
\cite{ger:Das}
\begin{gather*}
\frac{d\phi^{a}(x)}{dt_{01}}=\{\phi^{a}(x),H_{0}\}_{1}=\int_{0}^{2\pi}
P_{1}^{ab}(x,y)\frac{\partial h_{0}}{\partial \phi^{b}(y)}dy =
\int_{0}^{2\pi}R^{a}_{c}(x,z)P_{0}^{cb}(z,y) \frac{\partial
h_{0}}{\partial \phi^{b}(y)}dy,\\
\frac{d\phi^{a}(x)}{dt_{10}}=\{\phi^{a}(x),H_{1}\}_{0}=\int_{0}^{2\pi}
P_{0}^{ab}(x,y)\frac{\partial h_{1}}{\partial
\phi^{b}(y)}dy=\int_{0}^{2\pi}(R^{-1})^{a}_{c}(x,z)P_{0}^{cb}(z,y)
\frac{\partial h_{0}} {\partial \phi^{b}(y)}dy.
\end{gather*}

 There $R^{a}_{b}(x,y)$ is a recursion operator and $(R^{-1})^{a}_{b}(x,y)$
is its inverse
\begin{gather*}%\label{eq3}
R^{a}_{c}(x,y)=\int _{0}^{2\pi}P^{ab}_{1}
(x,z)(P_{0})^{-1}_{bc}(z,y)dz.
\end{gather*}
The first branch of the hierarchies of dynamical systems has the
following form
\begin{gather*}\frac{d\phi^{a}(x)}{dt_{0N}}=\int_{0}^{2\pi}\!\!
(R(x,y_{1}){\cdots} R(y_{N-1}))^{a}_{c}
P_{0}^{cb}(y_{N-1},y_{N})\frac{\partial h_{0}}{\partial
\phi^{b}(y_{N})}dy_{1}{\cdots} dy_{N},\quad  N=1,2,\dots,\infty.
\end{gather*}
 The second branch of the hierarchies can be obtained by replacement $R \to R^{-1}$ and
$t_{0N} \to t_{N0}$. We will consider only the f\/irst branch of
the hierarchies.

 The local PBs of hydrodynamic type  were introduced by Dubrovin,
Novikov \cite{ger:DN,ger:DNs} for Hamiltonian description of
equations of hydrodynamics. They were generalized by Ferapontov
\cite{ger:Fer} and Mokhov, Ferapontov \cite{ger:Mok4} to the
non-local PBs of hydrodynamic type. The hydrodynamic type systems
were considered by Tsarev \cite{ger:Tsa}, Maltsev \cite{ger:Mal},
Ferapontov \cite{ger:Fer2}, Mokhov \cite{ger:Mok3} (see
also~\cite{ger:Mok0}), Pavlov~\cite{ger:Pav} (see
also~\cite{ger:Pav1}), Maltsev, Novikov~\cite{ger:Nov}. The
polynomials of local chiral currents were considered by Goldshmidt
and Witten \cite{ger:Gol} (see also \cite{ger:Are}). The local
conserved chiral charges in principal chiral models were
considered by Evans, Hassan, MacKay, Mountain \cite{ger:Eva}. The
tensor nonlocal chiral charges were introduced by Pohlmeyer
\cite{ger:Pol} (see also \cite{ger:Meu,ger:Thi}). The string
models of hydrodynamic type were considered by author
\cite{ger:Ger,ger:Ger1}.
 In Section~3, the author applied hydrodynamic approach to integrable
systems to obtain new integrable string equations. In Section~4,
the author used the nonlocal Pohlmeyer charges to obtain a new
string equation in terms of the nonlocal currents. In Section~5,
the author applied the local invariant chiral currents to a simple
Lie algebra to construct new integrable string equations.

\section{String model of principal chiral model type}

 A string model is described by the Lagrangian
\begin{gather}\label{eq4}L=\frac{1}{2}\int_{0}^{2\pi}\eta^{\alpha\beta}g_{ab}
(\phi (t,x))\frac{\partial \phi ^{a}(t,x)} {\partial
x^{\alpha}}\frac{\partial \phi ^{b}(t,x)}{\partial
x^{\beta}}dx\end{gather} and by two f\/irst kind constraints
\begin{gather}g_{ab}(\phi (x))\left[\frac{\partial\phi^{a}(x)}{\partial t}
\frac{\partial\phi^{b}(x)}{\partial
t}+\frac{\partial\phi^{a}(x)}{\partial
x}\frac{\partial\phi^{b}(x)} {\partial x}\right]\approx 0,\qquad
g_{ab}(\phi (x))\frac{\partial\phi^{a}(x)}{\partial t}
\frac{\partial\phi^{b}(x)}{\partial x}\approx
0.\nonumber\end{gather} The target space local coordinates
$\phi^{a}(x)$, $a=1,\dots,n$ belong to certain given smooth
$n$-dimen\-sional manifold $M^{n}$ with nondegenerate metric
tensor
\begin{gather*}%\label{eq5}
g_{ab}(\phi(x))=\eta_{\mu\nu}e_{a}^{\mu}(\phi(x))
e_{b}^{\nu}(\phi(x)),
\end{gather*} where $\mu ,\nu
=1,\dots ,n$ are indices of tangent space to manifold $M^{n}$ on
some point $\phi^{a}(x)$. The veilbein $e_{a}^{\mu}(\phi)$ and its
inverse $e_{\mu}^{a}(\phi)$ satisfy the conditions
\begin{gather}e_{a}^{\mu}e_{\mu}^{b}=\delta _{a}^{b},\qquad e_{a}^{\mu}e^{a\nu }
=\eta ^{\mu\nu}. \nonumber\end{gather} The coordinates
$x^{\alpha}$ $(x^{0}=t$, $x^{1}=x)$ belong to world sheet with
metric tensor $g_{\alpha\beta}$ in conformal gauge. The string
equations of motion have the form
\begin{gather*}%\label{eq6}
\eta^{\alpha\beta}[\partial _{\alpha\beta}\phi^{a}+ \Gamma
^{a}_{bc}(\phi)\partial_{\alpha }
\phi^{b}\partial_{\beta}\phi^{c}]=0,\qquad  \partial _{\alpha
}=\frac{\partial }{\partial x^{\alpha}},\qquad \alpha =0,1,
\end{gather*} where
\begin{gather*}%\label{eq7}
\Gamma^{a}_{bc}(\phi)=\frac{1}{2}e^{a}_{\mu}\left[\frac{\partial
e^{\mu}_{b}}{\partial \phi^{c}}+ \frac{\partial
e^{\mu}_{c}}{\partial \phi^{b}}\right]
\end{gather*}
is the connection. In terms of canonical currents
 \begin{gather*}%\label{eq8}
 J^{\mu}_{\alpha}(\phi)=e^{\mu}_{a}(\phi)\partial_{\alpha}\phi^{a}
 \end{gather*}
 the equations of motion have the form
\begin{gather*}\eta^{\alpha\beta}\partial_{\alpha}J^{\mu}_{\beta}(\phi(t,x))
=0,\qquad
\partial_{\alpha}J^{\mu}_{\beta}(\phi)-\partial_{\beta}J^{\mu}_{\alpha}(\phi)-
C^{\mu}_{\nu\lambda}(\phi)
J^{\nu}_{\alpha}(\phi)J^{\lambda}_{\beta}(\phi)=0,
\end{gather*}
where
\begin{gather*}%\label{eq9}
C^{\mu}_{\nu\lambda}(\phi)=e^{a}_{\nu}e^{b}_{\lambda}\left[\frac{\partial
e^{\mu}_{a}}{\partial \phi^{b}} -\frac{\partial
e^{\mu}_{b}}{\partial \phi^{a}}\right]
\end{gather*} is the torsion. The Hamiltonian has the
form
\begin{gather*}%\label{eq10}
H=\frac{1}{2}\int_{0}^{2\pi}[\eta^{\mu\nu}
J_{0\mu}J_{0\nu}+\eta_{\mu\nu} J^{\mu}_{1}J^{\nu}_{1}]dx,
\end{gather*} where
$J_{0\mu}(\phi)=e^{a}_{\mu}(\phi)p_{a}$,
$J^{\mu}_{1}(\phi)=e^{\mu}_{a}\frac{\partial }{\partial
x}\phi^{a}$ and
$p_{a}(t,x)=\eta_{\mu\nu}e^{\mu}_{a}e^{\nu}_{b}\frac{\partial}{\partial
t}\phi^{b}$ is the canonical momentum. The canonical commutation
relations of currents are as follows
\begin{gather*}
\{J_{0\mu}(\phi(x)),J_{0\nu}(\phi(y))\}=C^{\lambda}_{\mu\nu}(\phi(x))
J_{0\lambda}(\phi(x))
\delta (x-y),\\
\{J_{0\mu}(\phi(x)),J^{\nu}_{1}(\phi(y))\}=C^{\nu}_{\mu\lambda}(\phi(x))
J_{1}^{\lambda}(\phi(x))\delta(x-y)
-\frac{1}{2}\delta^{\nu}_{\mu}\frac{\partial}{\partial x}\delta
(x-y),
\\
\{J^{\mu}_{1}(\phi(x)),J^{\nu}_{1}(\phi(y))\}=0.
\end{gather*}
Let us introduce chiral currents
\[U^{\mu}=\eta^{\mu\nu}J_{0\nu}+J^{\mu}_{1},\qquad V^{\mu}=\eta^{\mu\nu}J_{0\nu}-
J^{\mu}_{1}\] The commutation relations of chiral currents are the
following
\begin{gather*}\{U^{\mu}(\phi(x)),U^{\nu}(\phi(y))\}=C^{\mu\nu}_{\lambda}(\phi(x))\left[\frac{3}{2}U^{\lambda}(\phi(x))-\frac{1}{2}
V^{\lambda}(\phi(x))\right]\delta (x-y)
-\eta^{\mu\nu}\frac{\partial}{\partial x}\delta
(x-y),\nonumber\\
\{U^{\mu}(\phi(x)),V^{\nu}(\phi(y))\}=C^{\mu\nu}_{\lambda}(\phi(x))
[U^{\lambda}(\phi(x))+V^{\lambda}(\phi(x))]\delta
(x-y),\\ %\label{eq11}\\
\{V^{\mu}(\phi(x)),V^{\nu}(\phi(y))\}=C^{\mu\nu}_{\lambda}(\phi(x))\left
[\frac{3}{2}V^{\lambda}(\phi(x))-\frac{1}{2}U^{\lambda}(\phi(x))\right]\delta
(x-y) +\eta^{\mu\nu}\frac{\partial}{\partial x}\delta (x-y).
\nonumber
\end{gather*} Equations of motion in light-cone
coordinates
\[x^{\pm}=\frac{1}{2}(t\pm x),\qquad \frac{\partial }{\partial x^{\pm }}=
\frac{\partial}{\partial t}\pm \frac{\partial}{\partial x}\] have
the form
\begin{gather*}%\label{eq12}
\partial _{-}U^{\mu}=C^{\mu}_{\nu\lambda}(\phi(x))U^{\nu}
V^{\lambda},\qquad \partial _{-}V^{\mu}=
C^{\mu}_{\nu\lambda}(\phi(x))V^{\nu}U^{\lambda}.\end{gather*} In
the case of the null torsion
\begin{gather}C^{\mu}_{\nu\lambda}=0,\qquad e^{\mu}_{a}(\phi)=
\frac{\partial e^{\mu}}{\partial \phi^{a}},\qquad \Gamma
^{a}_{bc}(\phi)=e^{a}_{\mu}\frac{\partial ^{2}e^{\mu}}{{\partial
\phi^{b}}{\partial \phi^{c}}},\qquad
R^{\mu}_{\nu\lambda\rho}(\phi)=0\nonumber
\end{gather}
the string model is integrable.The Hamiltonian equations of motion
under Hamiltonian (\ref{eq4}) are described by two independent
left and right movers: $U^{\mu}(t+x)$ and $V^{\mu}(t-x)$.

\section{Integrable string models of hydrodynamic type\\ with  null torsion}

 We want to construct new integrable string models with
Hamiltonians as polynomials of the initial chiral currents
$U^{\mu}(\phi(x))$. The PB of chiral currents $U^{\mu}(x)$
coincides with the f\/lat PB of Dubrovin, Novikov
\begin{gather*}%\label{eq13}
\{U^{\mu}(x),U^{\nu}(y)\}_{0}=-\eta^{\mu\nu}\frac{\partial }
{\partial x}\delta (x-y).
\end{gather*} Let us introduce a local Dubrovin, Novikov
PB \cite{ger:DN,ger:DNs}. It has the form
\begin{gather*}%\label{eq14}
\{U^{\mu}(x),U^{\nu}(y)\}_{1}=g^{\mu\nu}(U(x))\frac{\partial }
{\partial x}\delta (x-y)-
\Gamma^{\mu\nu}_{\lambda}(U(x))\frac{\partial
U^{\lambda}(x)}{\partial x}\delta (x-y).\end{gather*} This PB is
skew-symmetric if $g^{\mu\nu}(U)=g^{\nu\mu}(U)$ and it satisf\/ies
Jacobi identity if $\Gamma^{a}_{bc}(U)=\Gamma^{a}_{cb}(U)$,
$C^{a}_{bc}(U)=0$, $R^{a}_{bcd}(U)=0.$ In the case of non-zero
curvature tensor we must include Weingarten operators into right
side of the PB with the step-function ${\rm sgn}\,
(x-y)$=$(\frac{d}{dx})^{-1}\delta (x-y)$=$\nu (x-y)$
\cite{ger:Fer,ger:Mok4}. The PBs $\{*,*\}_{0}$ and $\{*,*\}_{1}$
are compatible by Magri \cite{ger:Mag} if the pencil
$\{*,*\}_{0}+\lambda \{*,*\}_{1}$ is also PB . As a result, Mokhov
\cite{ger:Mok3,ger:Mok2} obtained the compatible pair of PBs
\begin{gather*}
P_{0\mu\nu}(U)(x,y)=-\eta_{\mu\nu}\frac{\partial }
{\partial x}\delta (x-y),\nonumber\\
%\label{eq15}
P_{1\mu\nu}(U)(x,y)=2\frac{\partial^{2} F(U)} {\partial
U^{\mu}\partial U^{\nu}}\frac{\partial } {\partial x}\delta
(x-y)+\frac{\partial^{3} F(U)}{\partial U^{\mu}\partial
U^{\nu}\partial U^{\lambda}} \frac{\partial U^{\lambda}}{\partial
x}\delta (x-y).
\end{gather*}
The function $F(U)$ satisf\/ies the equation
\begin{gather*}%\label{eq16}
\frac{\partial^{3} F(U)}{\partial U^{\mu}\partial U^{\rho}\partial
U^{\lambda}}\eta^{\lambda\rho}\frac{\partial^{3} F(U)} {\partial
U^{\nu}\partial U^{\sigma}\partial
U^{\rho}}=\frac{\partial^{3}F(U)} {\partial U^{\nu}\partial
U^{\rho}\partial U^{\lambda}}\eta^{\lambda\rho}\frac{\partial^{3}
F(U)} {\partial U^{\mu}\partial U^{\sigma}\partial U^{\rho}}.
\end{gather*} This equation is WDVV
\cite{ger:DW,ger:DVV} associativity equation and it was obtained
in 2D topological f\/ield theory. Dubrovin \cite{ger:Dub,ger:DZ}
obtained a lot of solutions of WDVV equation. He showed that local
f\/ields $U^{\mu}(x)$ must belong
 Frobenius manifolds to solve the WDVV equation and gave examples of
Frobenius structures. Associative Frobenius algebra may be written
in the following form
\begin{gather*}
\frac{\partial }{\partial U^{\mu}} * \frac{\partial } {\partial
U^{\nu}} := d_{\mu\nu}^{\lambda}(U)\frac{\partial }{\partial
U^{\lambda}}.\nonumber
\end{gather*} Totally symmetric structure
function has the form
\begin{gather*}d_{\mu\nu\lambda}(U)=\frac{\partial F(U)}{\partial U^{\mu}
\partial U^{\nu}
\partial U^{\lambda}},\qquad \mu,\nu,\lambda =1,\dots ,n\nonumber
\end{gather*}
and associativity condition
\begin{gather*}\left(\frac{\partial }{\partial U^{\mu}} * \frac{\partial }
{\partial U^{\nu}}\right) * \frac{\partial }{\partial U^{\lambda}}
= \frac{\partial }{\partial U^{\mu}} * \left(\frac{\partial
}{\partial U^{\nu}} *  \frac{\partial }{\partial
U^{\lambda}}\right)\nonumber
\end{gather*} leads to the WDVV equation.
Function $F(U)$ is quasihomogeneous function of its variables
\begin{gather*}%\label{eq17}
\left(d_{\mu}U^{\mu}\frac{\partial} {\partial U^{\mu}}\right)F(U)
=d_{F}F(U)+
A_{\mu\nu}U^{\mu}U^{\nu}+B_{\mu}U^{\mu}+C,\end{gather*} here
numbers $d_{\mu}$, $d_{F}$, $A_{\mu\nu}$, $B_{\mu}$, $C$ depend on
the type of polynomial function $F(U)$. Here are some Dubrovin
examples of solutions of the WDVV equation
\begin{gather} n=1,\qquad F(U)=U^{3}_{1};\nonumber\\
 \label{eq18}n=2,\qquad F(U)= \frac{1}{2}U^{2}_{1}U_{2}+e^{U_{2}} ,\qquad d_{1}=1,\qquad
d_{2}=2, \qquad d_{F}=2, \qquad A_{11}=1
\end{gather}
and quasihomogeneity condition for $n=2$ has the form
\[\left(d_{1}U_{1}\frac{\partial }{\partial U_{1}}+d_{2}\frac{\partial }
{\partial U_{2}}\right)F(U)=d_{F}F(U)+A_{11}U^{2}_{1}.\]

 We used local f\/ields $U_{\mu}$ with low indices here for
convenience. One of the Dubrovin polynomial solutions is
\begin{gather}\label{eq19}F(U)=\frac{1}{2}(U^{2}_{1}U_{3}+U_{1}U^{2}_{2})+
\frac{1}{4}U^{2}_{2}U^{2}_{3}+ \frac{1}{60}U^{5}_{3} ,\end{gather}
here $d_{1}=1$, $d_{2}=\frac{3}{2}$, $ d_{3}=2$, $d_{F}=4$ and the
polynomial function
\[f(U_{2},U_{3})=\frac{1}{4}U^{2}_{2}U^{2}_{3}+\frac{1}{60}U^{5}_{3}\]
is a solution of the additional PDE.

 In the bi-Hamiltonian approach to an integrable string model we must construct
the recursion operator to generate a hierarchy of PBs and a
hierarchy of Hamiltonians
\begin{gather*}
R^{\mu}_{\nu}(x,y)=\int_{0}^{2\pi}P^{\mu\lambda}_{1}(x,z)
(P^{-1}_{0}(z,y))_{\lambda\nu}dz\nonumber\\
\phantom{R^{\mu}_{\nu}(x,y)}{}
=2\frac{\partial^{2}F(U(x))}{\partial U^{\mu}(x)\partial
U^{\nu}(x)}\delta (x-y)+\frac{\partial^{3} F(U(x))}{\partial
U^{\mu}(x)\partial U^{\nu}(x)
\partial U^{\lambda}(x)}\,\frac{\partial U^{\lambda}(y)}{\partial y}\nu(x-y).%\label{eq20}
\end{gather*}

 The Hamiltonian equation of motion with Hamiltonian $H_{0}$ is the following
\begin{gather*}%\label{eq21}
H_{0}=\int_{0}^{2\pi}\eta_{\mu\nu}U^{\mu}(x) U^{\nu}(x)dx,\qquad
\frac{\partial U^{\mu}}{\partial t}= \frac{\partial
U^{\mu}}{\partial x}.\end{gather*} First of the new equations of
motion under the new time $t_{1}$ has the form \cite{ger:Mok3}
\begin{gather}\label{eq22}\frac{\partial U^{\mu}}{\partial t_{1}}=
\int_{0}^{2\pi}R^{\mu}_{\nu}(x,y)\frac{\partial U^{\mu}(y)}
{\partial y}dy=\eta^{\mu\nu}\frac{d}{dx}\left(\frac{\partial
F(x)}{\partial U^{\nu}}\right).\end{gather} This equation of
motion can be obtained as Hamiltonian equation with new
Hamiltonian $H_{1}$
\begin{gather*}%\label{eq23}
H_{1}=\int_{0}^{2\pi}\frac{\partial F(U(x))} {\partial
U^{\mu}}U^{\mu}(x)dx,\end{gather*} where $F(U)$ is each of
Dubrovin solutions WDVV associativity equation (\ref{eq18}),
(\ref{eq19}). Any system of the following hierarchy
\cite{ger:Mok3}
\begin{gather*}%\label{eq24}
\frac{\partial U^{\mu}}{\partial t_{M}}=\int_{0}^{2\pi}
(R(x,y_{1})\cdots R(y_{M-1},y_{M}))^{\mu}_{\nu} \frac{\partial
U^{\nu}}{\partial y_{M}}dy_{1}\cdots dy_{M}\end{gather*} is an
integrable system. As result we obtain chiral currents
$U^{\mu}(\phi(t_{M},x))=f^{\mu}(\phi(t_{M},x)$, where
$f^{\mu}(\phi)$ is a solution of the equation of motion. In the
case of the Hamiltonian $H_{1}$ and of the equation of
motion~(\ref{eq22}) we can introduce new currents
\begin{gather*}%\label{eq25}
J^{\mu}_{0}(t_{1},x)=U^{\mu}(t_{1},x),\,\,J^{\mu}_{1}(t_{1},x)=
\eta^{\mu\nu}\frac{\partial F(U(t_{1},x))} {\partial
U^{\nu}}.\end{gather*}
 Consequently, we can introduce a new metric tensor and a new veilbein depending of the
 new time coordinate. The equation for the new
 metric tensor has the form
\begin{gather*}%\label{eq26}
e^{\mu}_{a}(\phi(t_{1},x))\frac{\partial \phi^{a}(t_{1},x)}
{\partial x}=\frac{de^{\mu}(\phi(t_{1},x))}{dx} =\eta
^{\mu\nu}\frac{\partial F(f(\phi(t_{1},x)))}{\partial
f^{\nu}(\phi(t_{1},x))}.\end{gather*}

\section{New string equation in terms\\ of Pohlmeyer tensor
nonlocal currents}

  In the case of the f\/lat space
$C^{\mu}_{\nu\lambda}=0$ there exist nonlocal totally symmetric
tensor chiral currents called ``Pohlmeyer'' currents
\cite{ger:Pol,ger:Meu, ger:Thi}
\begin{gather*}R^{(M)}(U(x))\equiv R^{(\mu _{1}\mu _{2}\dots \mu _{M})}(U(x))\nonumber\\
\phantom{R^{(M)}(U(x))}{} = U^{(\mu _{1}}(x) \int_{0}^{x}U^{\mu
_{2}}(x_{1})dx_{1}\cdots \int_{0}^{x_{M-2}}U^{\mu_{M})}(x_{M-1})dx_{M-1},%\label{eq27}
\end{gather*}
where round brackets the mean totally symmetric product of chiral
currents $U^{\mu}(U)$. The new Hamiltonians may have the following
forms
\begin{gather*}%\label{eq28}
H^{(M)}=\frac{1}{2}\int_{0}^{2\pi}R^{(M)}(U(x))d_{2M}
R^{(M)}(U(x))dx,\end{gather*} where $d_{M}\equiv d_{(\mu _{1}\mu
_{2}\dots \mu _{M})}$ is totally symmetric invariant constant
tensor, which can be constructed from Kronecker deltas. For
example
\begin{gather*}%\label{eq29}
R^{(2)}\equiv R^{\mu\nu}(U(x))=\frac{1}{2}[U^{\mu}(x)\int
\limits_{0}^{x}U^{\nu}(x_{1})dx_{1}+U^{\nu}(x)
\int_{0}^{x}U^{\mu}(x_{1})dx_{1}],\\
H^{(2)}=\frac{1}{2}\int_{0}^{2\pi}\Bigg[U^{\mu}(x)U^{\mu}(x)
\int_{0}^{x}U^{\nu}(x_{1})dx_{1}
\int_{0}^{x}U^{\nu}(x_{2})dx_{2}\nonumber\\
\phantom{H^{(2)}=}{}
+U^{\mu}(x)U^{\nu}(x)\int_{0}^{x}U^{\mu}(x_{1})dx_{1}\int_{0}^{x}
U^{\nu}(x_{2})dx_{2}\Bigg]dx.%\label{eq30}
\end{gather*}
The Hamiltonian $H^{(2)}$ commutes with the Hamiltonian
$H^{(1)}=\frac{1}{2}\int_{0}^{2\pi}U^{\mu}(x)U^{\mu}(x)dx$ and it
commutes with the Casimir $\int_{0}^{2\pi}U^{\mu}(x)dx$. The
equation of motion under the Hamiltonian $H^{(2)}$ is as follows
\begin{gather*}\frac{\partial U^{\mu}(x)}{\partial t}=\frac{\partial }
{\partial x}\Bigg[U^{\mu}(x)\int_{0}^{x}\!
U^{\nu}(x_{1})dx_{1}\int_{0}^{x}\! U^{\nu}(x_{2})dx_{2}+
U^{\nu}(x)\int_{0}^{x} \! U^{\mu}(x_{1})dx_{1}
\int_{0}^{x}\! U^{\nu}(x_{2})dx_{2}\Bigg]\nonumber\\
\phantom{\frac{\partial U^{\mu}(x)}{\partial t}=}{}
-U^{\nu}(x)U^{\nu}(x)\int_{0}^{x}U^{\mu}(x_{1})dx_{1}-U^{\mu}(x)
U^{\nu}(x)\int_{0}^{x}U^{\nu}(x_{1})dx_{1}.%\label{eq31}
\end{gather*} In the variables
\begin{gather*}%\label{eq32}
S^{\mu}(x)=\int_{0}^{x}U^{\mu}(y)dy\end{gather*} the latter
equation can be rewritten as follows
\begin{gather*}%\label{eq33}
\frac{\partial S^{\mu}}{\partial t}=\frac{\partial }{\partial x}
(S^{\mu}(S^{\nu}S^{\nu}))+
\int_{0}^{x}S^{\mu}\left(S^{\nu}\frac{\partial
^{2}S^{\nu}}{\partial ^{2}y}\right)dy,\qquad  \mu, \nu =1,2,\dots
,n.\end{gather*}

\section{Integrable string models with constant torsion}

  Let us go back to the commutation relations of chiral currents.
Let the torsion $C^{\mu}_{\nu\lambda}(\phi(x))\ne 0$ and
$C_{\mu\nu\lambda}=f_{\mu\nu\lambda}$ be structure constant of s
simple Lie algebra. We will consider a string model with the
constant torsion in light-cone gauge in target space. This model
coincides with the principal chiral model on compact simple Lie
group. We cannot divide the motion on right and left mover because
of chiral currents
$\partial_{-}U^{\mu}=f^{\mu}_{\nu\lambda}U^{\nu}V^{\lambda}$,
$\partial_{-}V^{\mu}= f^{\mu}_{\nu\lambda}V^{\nu}U^{\lambda}$ are
not conserved. The correspondent charges are not Casimirs. The
present paper was stimulated by paper~\cite{ger:Eva}. Evans,
Hassan, MacKay, Mountain (see \cite{ger:Eva} and references
therein) constructed local invariant chiral currents as
polynomials of the initial chiral currents of $SU(n)$, $SO(n)$,
$SP(n)$ principal chiral models and they found such combination of
them that the corresponding charges are Casimir operators of these
dynamical systems. Their paper was based on the paper of de
Azcarraga, Macfarlane, MacKay, Perez Bueno (see \cite{ger:Az} and
references therein) about invariant tensors for simple Lie
algebras. Let $t_{\mu}$ be $n \otimes n$ traceless hermitian
matrix representations of generators Lie algebra
\begin{gather*}%\label{eq34}
[t_{\mu},t_{\nu}]=2if_{\mu\nu\lambda}t_{\lambda} ,\qquad {\rm Tr}
(t_{\mu}t_{\nu})=2\delta_{\mu\nu}.\end{gather*} Here is an
additional relation for $SU(n)$ algebra
\begin{gather*}%\label{eq35}
\{t_{\mu},t_{\nu}\}=\frac{4}{n}\delta_{\mu\nu}+ 2d_{\mu\nu\lambda}
t_{\lambda}, \qquad \mu=1,\dots ,n^{2}-1 .\end{gather*} De
Azcarraga et al.\ gave some examples of invariant tensors of
simple Lie algebras and they gave a general method to calculate
them. Invariant tensors may be constructed as invariant symmetric
polynomials on $SU(n)$
\[d^{(M)}_{(\mu_{1}\dots \mu_{M})}= \frac
{1}{M!}{\rm STr}(t_{\mu_{1}}\cdots t_{\mu_{M}}),\] where ${\rm
STr}$ means the completely symmetrized product of matrices and
$d^{(M)}_{(\mu_{1}\dots \mu_{M})}$ is the totally symmetric tensor
and $M=2,3,\dots ,\infty $. Another family of invariant symmetric
tensors \cite{ger:Kle,ger:Sud} (see also \cite{ger:Az}) called
$D$-family based on the product of the symmetric structure
constant $d_{\mu\nu\lambda}$ of the $SU(n)$ algebra is as follows:
\begin{gather*}D^{(M)}_{(\mu_{1}\dots \mu_{M})}=
d^{k_{1}}_{({\mu_{1}\mu_{2}}}d^{k_{1}k_{2}}_{\mu_{3}}\cdots
d^{k_{M-2}
k_{M-3}}_{\mu_{M-2}}d^{k_{M-3}}_{\mu_{M-1}\mu_{M})},\nonumber
\end{gather*}
where
$D^{(2)}_{\mu\nu}=\delta_{\mu\nu}$,\,$D^{(3)}_{\mu\nu\lambda}=d_{\mu\nu\lambda}$
and $M=4,5,\dots ,\infty.$

Here are $n-1$ primitive invariant tensors on $SU(n)$. The
invariant tensors for $M \ge n$ are functions of primitive
tensors. The Casimir operators on $SU(n)$ algebra have the form
\begin{gather*}
C^{(M)}(t)=d^{M}_{(\mu_{1}\dots \mu_{M})}t_{\mu_{1}}\cdots
t_{\mu_{M}}. \nonumber\end{gather*} Evans et al.\ introduced local
chiral currents based on the invariant symmetric polynomials on
simple Lie groups
\begin{gather}\label{eq36}J^{(M)}(U)={\rm STr} (U\cdots U)\equiv {\rm STr}\,U^{M}=
d^{(M)}_{\mu_{1}\dots \mu_{M}}U^{\mu_{1}}\cdots
U^{\mu_{M}},\end{gather} where $U=t_{\mu}U^{\mu}$ and $\mu
=1,\dots ,n^{2}-1$ . It is possible to decompose the invariant
symmetric chiral currents $J^{(M)}(U)$ into product of the basic
invariant chiral currents $D^{(M)}(U)$
\begin{gather*}%\label{eq37}
D^{(2)}(U)=d^{(2)}_{\mu\nu}U^{\mu}U^{\nu}=\eta_{\mu\nu}U^{\mu}U^{\nu},\qquad
D(3)(U)= d_{\mu\nu\lambda}U^{\mu}U^{\nu}U^{\lambda},
\\
 D^{(M)}(U(x))=
d^{k_{1}}_{\mu_{1}\mu_{2}}d^{k_{1}k_{2}}_{\mu_{3}}\cdots
d^{k_{M-2}k_{M-3}}_{\mu_{M-2}} d^{k_{M-3}}_{\mu_{M-1}\mu_{M}}
U_{\mu_{1}}U_{\mu_{2}}\cdots U_{\mu_{M}},\nonumber
\end{gather*}
where $M=4,5,\dots ,\infty$.
 The author obtained the following expressions for local invariant chiral currents $J^{(M)}(U)$
\begin{gather*}J^{(2)}=2D^{(2)},  \qquad J^{(3)}=2D^{(3)},\qquad
J^{(4)}=2D^{(4)}+\frac{4}{n}D^{(2)2},\\
J^{(5)}=2D^{(5)}+\frac{8}{n}D^{(2)}D^{(3)},\qquad
%\label{eq38}
J^{(6)}=2D^{(6)}+\frac{4}{n}D^{(3)2}+\frac{8}{n}
D^{(2)}D^{(4)}+ \frac{8}{n^{2}}D^{(2)3},\\
J^{(7)}=2D^{(7)}+\frac{8}{n}D^{(3)}D^{(4)}+\frac{8}{n}D^{(2)}D^{(5)}+
\frac{24}{n^{2}}D^{(2)2}D^{(3)},\nonumber\\
J^{(8)}=2D^{(8)}+\frac{4}{n}D^{(4)2}\!+\frac{8}{n}D^{(3)}D^{(5)}\!+
\frac{8}{n}D^{(2)}D^{(6)}\!+\frac{24}{n^{2}}D^{(2)}D^{(3)2}\!+\frac{24}{n^{2}}D^{(2)2}D^{(4)}\!+\frac{16}
{n^{3}}D^{(2)4},\nonumber\\
J^{(9)}=2D^{(9)}+\frac{8}{n}D^{(4)}D^{(5)}+\frac{8}{n}D^{(3)}D^{(6)}+\frac{8}{n}D^{(2)}D^{(7)}
+\frac{8}{n^{2}}D^{(3)3}+\frac{48}{n^{2}}D^{(2)}D^{(3)}D^{(4)}\nonumber\\
\phantom{J^{(9)}=}{}
+\frac{24}{n^{2}}D^{(2)2}D^{(5)}+\frac{64}{n^{3}}D^{(2)3}D^{(3)}.\nonumber
\end{gather*}
Both families of invariant chiral currents $J^{(M)}(U(x))$ and
$D^{(M)}(U(x))$ satisfy the conservation equations $\partial_{-}
J^{(M)}(U(x))=0$, $\partial_{-} D^{(M)}(U(x))=0$.

 The commutation relations of invariant chiral currents $J^{(M)}(U(x))$ show that
these currents are not densities of dynamical Casimir operators
for $SU(n)$ group. Therefore, we will not consider these currents
in the following.

 We considered abasic family of invariant chiral currents
$D^{(M)}(U)$ and we proved that the invariant chiral currents
$D^{(M)}(U)$ form closed algebra under canonical PB and
corresponding charges are dynamical Casimir operators. The
commutation relations of invariant chiral currents $D^{(M)}(U(x))$
and $D^{(N)}(U(y))$ for $M,N = 2,3,4$ and for $M=2$, $N=2,3,\dots
,\infty$ are as follows
\begin{gather*}\{D^{(M)}(x),D^{(N)}(y)\}=-MN D^{(M+N-2)}(x)
\frac{\partial }{\partial x}\delta (x-y)\nonumber\\
\phantom{\{D^{(M)}(x),D^{(N)}(y)\}=}{} -
\frac{MN(N-1)}{M+N-2}\frac{\partial D^{(M+N-2)}(x)} {\partial
x}\delta(x-y).%\label{41}
\end{gather*} The commutation relations  for $M
\ge 5$, $N \ge 3$ are as follows
\begin{gather*}
\{D^{(5)}(x),D^{(3)}(y)\}=-[12D^{(6)}(x)+3D^{(6,1)}(x)]\frac{\partial
}{\partial x}
\delta (x-y) \nonumber\\
\phantom{\{D^{(5)}(x),D^{(3)}(y)\}=}{} -
\frac{1}{3}\frac{\partial}{\partial
x}[12D^{(6)}(x)+3D^{(6,1)}(x)]\delta (x-y),\nonumber\\
\{D^{(5)}(x),D^{(4)}(y)\}=-[16D^{(7)}(x)+4D^{(7,1)}(x)]\frac{\partial
}{\partial x}
\delta (x-y) \nonumber\\
\phantom{\{D^{(5)}(x),D^{(4)}(y)\}=}{} -
\frac{3}{7}\frac{\partial}{\partial
x}[16D^{(7)}(x)+4D^{(7,1)}(x)]\delta (x-y),\nonumber\\
\{D^{(6)}(x),D^{(3)}(y)\}=-[12D^{(7)}(x)+6D^{(7,1)}(x)]\frac{\partial
}{\partial x}
\delta (x-y) \nonumber\\
\phantom{\{D^{(6)}(x),D^{(3)}(y)\}=}{} -
\frac{2}{7}\frac{\partial}{\partial
x}[12D^{(7)}(x)+6D^{(7,1)}(x)]\delta (x-y),\nonumber\\
\{D^{(5)}(x),D^{(5)}(y)\}=-[16D^{(8)}(x)+8D^{(8,1)}(x)+D^{(8,2)}(x)]\frac{\partial
}
{\partial x}\delta (x-y)\nonumber\\
\phantom{\{D^{(5)}(x),D^{(5)}(y)\}=}{}
-\frac{1}{2}\frac{\partial}{\partial
x}[16D^{(8)}(x)+8D^{(8,1)}(x)+D^{(8,2)}(x)]
\delta (x-y),\nonumber\\
\{D^{(6)}(x),D^{(4)}(y)\}=-[16D^{(8)}(x)+8D^{(8,3)}(x)]\frac{\partial
}
{\partial x}\delta (x-y)\nonumber\\
\phantom{\{D^{(6)}(x),D^{(4)}(y)\}=}{}
-\frac{3}{8}\frac{\partial}{\partial
x}[16D^{(8)}(x)+8D^{(8,3)}(x)] \delta (x-y),\nonumber\\
\{D^{(7)}(x),D^{(3)}(y)\}=-[12D^{(8)}(x)+6D^{(8,1)}(x)+3D^{(8,3)}(x)]\frac{\partial
}
{\partial x}\delta (x-y)\nonumber\\
\phantom{\{D^{(7)}(x),D^{(3)}(y)\}=}{}
-\frac{1}{4}\frac{\partial}{\partial
x}[12D^{(8)}(x)+6D^{(8,1)}(x)+3D^{(8,3)}(x)]
\delta (x-y),\nonumber\\
\{D^{(8)}(x),D^{(3)}(y)\}=-[12D^{(9)}(x)+6D^{(9,1)}(x)+6D^{(9,2)}(x)]\frac{\partial
}
{\partial x}\delta (x-y)\nonumber\\
\phantom{\{D^{(8)}(x),D^{(3)}(y)\}=}{}
-\frac{2}{9}\frac{\partial}{\partial
x}[12D^{(9)}(x)+6D^{(9,1)}(x)+6D^{(9,2)}(x)]
\delta (x-y),\nonumber\\
\{D^{(7)}(x),D^{(4)}(y)\}=-[16D^{(9)}(x)+8D^{(9,2)}(x)+4D^{(9,3)}(x)]\frac{\partial
}
{\partial x}\delta (x-y)\nonumber\\
\phantom{\{D^{(7)}(x),D^{(4)}(y)\}=}{}
-\frac{1}{3}\frac{\partial}{\partial
x}[16D^{(9)}(x)+8D^{(9,2)}(x)+4D^{(9,3)}(x)]
\delta (x-y),\nonumber\\
\{D^{(6)}(x),D^{(5)(y)}\}=-[16D^{(9)}(x)+4D^{(9,1)}(x)+8D^{(9,2)}(x)+2D^{(9,4)}(x)]
\frac{\partial }{\partial x}\delta (x-y)\nonumber\\
\phantom{\{D^{(6)}(x),D^{(5)(y)}\}=}{}
-\frac{4}{9}\frac{\partial}{\partial
x}[16D^{(9)}(x)+4D^{(9,1)}(x)+8D^{(9,2)}(x)+
2D^{(9,4)}(x)]\delta (x-y),\nonumber\\
\{D^{(9)}(x),D^{(3)}(y)\}=-[12D^{(10)}(x)+6D^{(10,1)}(x)+6D^{(10,2)}(x)+3D^{(10,3)}(x)]
\frac{\partial }{\partial x}\delta (x-y)\nonumber\\
\phantom{\{D^{(9)}(x),D^{(3)}(y)\}=}{}
-\frac{1}{5}\frac{\partial}{\partial
x}[12D^{(10)}(x)+6D^{(10,1)}(x)+6D^{(10,2)}(x)+
3D^{(10,3)}(x)]\delta (x-y),\!\!\nonumber\\
\{D^{(8)}(x),D^{(4)}(y)\}=-[16D^{(10)}(x)+8D^{(10,2)}(x)+8D^{(10,4)}(x)]
\frac{\partial }{\partial x}\delta (x-y)\nonumber\\
\phantom{\{D^{(8)}(x),D^{(4)}(y)\}=}{}
-\frac{3}{10}\frac{\partial}{\partial
x}[16D^{(10)}(x)+8D^{(10,2)}(x)+
8D^{(10,4)}(x)]\delta (x-y),\nonumber\\
\{D^{(7)}(x),D^{(5)}(y)\}=-[16D^{(10)}(x)+8D^{(10,3)}(x)+4D^{(10,1)}(x)\nonumber\\
\phantom{\{D^{(7)}(x),D^{(5)}(y)\}=}{} + 4D^{(10,4)}(x)+
2D^{(10,5)}(x)+D^{(10,6)}(x)] \frac{\partial }{\partial x}\delta
(x-y)\nonumber\\
\phantom{\{D^{(7)}(x),D^{(5)}(y)\}=}{}-\frac{2}{5}\frac{\partial}{\partial
x}[16D^{(10)}(x)+8D^{(10,3)}(x)+4D^{(10,1)}(x)+
4D^{(10,4)}(x)\nonumber\\
\phantom{\{D^{(7)}(x),D^{(5)}(y)\}=}{}+2D^{(10,5)}(x)+D^{(10,6)}(x)]\delta (x-y),\nonumber\\
\{D^{(6)}(x),D^{(6)}(y)\}=-[16D^{(10)}(x)+16D^{(10,2)}(x)+4D^{(10,7)}(x)]
\frac{\partial }{\partial x}\delta (x-y)\nonumber\\
\phantom{\{D^{(6)}(x),D^{(6)}(y)\}=}{}
-\frac{1}{2}\frac{\partial}{\partial
x}[16D^{(10)}(x)+16D^{(10,2)}(x)+
4D^{(10,7)}(x)]\delta (x-y).%\label{eq42}
\end{gather*}

 The new dependent invariant chiral currents $D^{(6,1)}$, $D^{(7,1)}$, $D^{(8,1)}- D^{(8,3)}$,
$D^{(9,1)}-D^{(9,4)}$, $D^{(10,1)}-D^{(10,7)}$ (see Appendix~A)
have the form
\begin{gather*}
D^{(6,1)}=d^{k}_{\mu\nu}d^{l}_{\lambda\rho}d^{n}_{\sigma\varphi}d^{kln}
U^{\mu}U^{\nu}U^{\lambda}U^{\rho}U^{\sigma}U^{\varphi},\nonumber\\
D^{(7,1)}=d^{k}_{\mu\nu}d^{l}_{\lambda\rho}d^{n}_{\sigma\varphi}d^{nm}_{\tau}
d^{klm}
U^{\mu}U^{\nu}U^{\lambda}U^{\rho}U^{\sigma}U^{\varphi}U^{\tau},\nonumber\\
D^{(8,1)}=[d^{k}_{\mu\nu}d^{kl}_{\lambda}d^{ln}_{\rho}][d^{m}_{\sigma\varphi}]
[d^{p}_{\tau\theta}]d^{nmp}
U^{\mu}U^{\nu}U^{\lambda}U^{\rho}U^{\sigma}U^{\varphi}U^{\tau}U^{\theta},\nonumber\\
D^{(8,2)}=[d^{k}_{\mu\nu}][d^{l}_{\lambda\rho}][d^{n}_{\sigma\varphi}]
[d^{m}_{\tau\theta}]d^{klp}d^{nmp}
U^{\mu}U^{\nu}U^{\lambda}U^{\rho}U^{\sigma}U^{\varphi}U^{\tau}U^{\theta},\nonumber\\
D^{(8,3)}=[d^{k}_{\mu\nu}d^{kl}_{\lambda}][d^{n}_{\rho\sigma}d^{nm}_{\varphi}]
[d^{p}_{\tau\theta}]d^{lmp}
U^{\mu}U^{\nu}U^{\lambda}U^{\rho}U^{\sigma}U^{\varphi}U^{\tau}U^{\theta},\nonumber\\
D^{(9,1)}=[d^{k}_{\mu\nu}d^{kl}_{\lambda}d^{ln}_{\rho}d^{nm}_{\sigma}]
[d^{p}_{\varphi\tau}][d^{r}_{\theta\omega}]d^{mpr}
U^{\mu}U^{\nu}U^{\lambda}U^{\rho}U^{\sigma}U^{\varphi}U^{\tau}U^{\theta}
U^{\omega},\nonumber\\
D^{(9,2)}=[d^{k}_{\mu\nu}d^{kl}_{\lambda}d^{ln}_{\rho}][d^{m}_{\sigma\varphi}
d^{mp}_{\tau}][d^{r}_{\theta\omega}]d^{npr}
U^{\mu}U^{\nu}U^{\lambda}U^{\rho}U^{\sigma}U^{\varphi}U^{\tau}U^{\theta}
U^{\omega},\nonumber\\
D^{(9,3)}=[d^{k}_{\mu\nu}d^{kl}_{\lambda}][d^{n}_{\rho\sigma}d^{nm}_{\varphi}]
[d^{p}_{\tau\theta}d^{pr}_{\omega}]d^{lmr}
U^{\mu}U^{\nu}U^{\lambda}U^{\rho}U^{\sigma}U^{\varphi}U^{\tau}U^{\theta}
U^{\omega},\nonumber\\
D^{(9,4)}=[d^{k}_{\mu\nu}d^{kl}_{\lambda}][d^{n}_{\rho\sigma}]
[d^{m}_{\varphi\tau}][d^{p}_{\theta\omega}]d^{lnr}d^{mpr}
U^{\mu}U^{\nu}U^{\lambda}U^{\rho}U^{\sigma}U^{\varphi}U^{\tau}U^{\theta}
U^{\omega},\nonumber\\
D^{(10,1)}=[d^{k}_{\mu\nu}d^{kl}_{\lambda}d^{ln}_{\rho}d^{nm}_{\sigma}
d^{mp}_{\varphi}][d^{r}_{\tau\theta}][d^{s}_{\omega\beta}]d^{prs}
U^{\mu}U^{\nu}U^{\lambda}U^{\rho}U^{\sigma}U^{\varphi}U^{\tau}U^{\theta}
U^{\omega}U^{\beta},\nonumber\\
D^{(10,2)}=[d^{k}_{\mu\nu}d^{kl}_{\lambda}d^{ln}_{\rho}d^{nm}_{\sigma}]
[d^{p}_{\varphi\tau}d^{pr}_{\theta}][d^{s}_{\omega\beta}]d^{mrs}
U^{\mu}U^{\nu}U^{\lambda}U^{\rho}U^{\sigma}U^{\varphi}U^{\tau}U^{\theta}
U^{\omega}U^{\beta},\nonumber\\
D^{(10,3)}=[d^{k}_{\mu\nu}d^{kl}_{\lambda}d^{ln}_{\rho}][d^{m}_{\sigma\varphi}
d^{mp}_{\varphi}d^{pr}_{\tau}][d^{s}_{\omega\beta}]d^{nrs}
U^{\mu}U^{\nu}U^{\lambda}U^{\rho}U^{\sigma}U^{\varphi}U^{\tau}U^{\theta}
U^{\omega}U^{\beta},\nonumber\\
D^{(10,4)}=[d^{k}_{\mu\nu}d^{kl}_{\lambda}d^{ln}_{\rho}][d^{m}_{\sigma\varphi}
d^{mp}_{\tau}][d^{r}_{\theta\omega}d^{rs}_{\beta}]d^{nps}
U^{\mu}U^{\nu}U^{\lambda}U^{\rho}U^{\sigma}U^{\varphi}U^{\tau}U^{\theta}
U^{\omega}U^{\beta},\nonumber\\
D^{(10,5)}=[d^{k}_{\mu\nu}d^{kl}_{\lambda}d^{ln}_{\rho}][d^{m}_{\sigma\varphi}]
[d^{p}_{\tau\theta}][d^{r}_{\omega\beta}]d^{nms}d^{prs}
U^{\mu}U^{\nu}U^{\lambda}U^{\rho}U^{\sigma}U^{\varphi}U^{\tau}U^{\theta}
U^{\omega}U^{\beta},\nonumber\\
D^{(10,6)}=[d^{k}_{\mu\nu}d^{kl}_{\lambda}][d^{n}_{\rho\sigma}d^{nm}_{\varphi}]
[d^{p}_{\tau\theta}][d^{r}_{\omega\beta}]d^{lms}d^{prs}
U^{\mu}U^{\nu}U^{\lambda}U^{\rho}U^{\sigma}U^{\varphi}U^{\tau}U^{\theta}
U^{\omega}U^{\beta},\nonumber\\
D^{(10,7)}=[d^{k}_{\mu\nu}d^{kl}_{\lambda}][d^{n}_{\rho\sigma}]d^{m}_{\varphi}]
[d^{mp}_{\tau\theta}][d^{r}_{\omega\beta}]d^{lns}d^{prs}
U^{\mu}U^{\nu}U^{\lambda}U^{\rho}U^{\sigma}U^{\varphi}U^{\tau}U^{\theta}
U^{\omega}U^{\beta}.
%\label{eq43}
\end{gather*}

 Let us apply the hydrodynamic approach to integrable string models
with constant torsion. In this case we must consider the conserved
primitive chiral currents $D^{(M)}(U(x))$, $(M=2,3,\dots ,n-1)$ as
local f\/ields of the Riemmann manifold. The non-primitive local
charges of invariant chiral currents with $M \ge n$ form the
hierarchy of new Hamiltonians in the bi-Hamiltonian approach to
integrable systems. The commutation relations of invariant chiral
currents are local PBs of hydrodynamic type.

 The invariant chiral currents $D^{(M)}$ with $M \ge 3$ for the
$SU(3)$ group can be obtained from the following relation
\begin{gather*}%\label{eq45}
d_{kln}d_{kmp}+d_{klm}d_{knp}+d_{klp}d_{knm}=
\frac{1}{3}(\delta_{ln}\delta_{mp}+\delta_{lm}\delta_{np}+\delta_{lp}
\delta_{nm}).\end{gather*}

 The corresponding invariant chiral currents for $SU(3)$ group have the
form
\begin{gather*}
%\label{46}
D^{(2N)}=\frac{1}{3^{N-1}}(\eta_{\mu\nu}U^{\mu}U^{\nu})^{N}=
\frac{1}{3^{N-1}}D^{(2)N},\\
%\label{eq47}
D^{(2N+1)}=\frac{1}{3^{N-1}}(\eta_{\mu\nu}U^{\mu}U^{\nu})^{N-1}d_{kln}U^{k}U^{l}U^{n}=
\frac{1}{3^{N-1}}D^{(2)N-1}D^{(3)}.\end{gather*} The invariant
chiral currents $D^{(2)}$, $D^{(3)}$ are local coordinates of the
Riemmann manifold $M^{2}$. The local charges $D^{(2N)}$, $N \ge 2$
form a hierarchy of Hamiltonians. The new nonlinear equations of
motion for chiral currents are as follows
\begin{gather*}\frac{\partial D^{(k)}(U(x))}{\partial
t_{N}}=\left\{D^{(k)}(U(x)),\int_{0}^{2\pi}D^{(2)N}(U(y))dy\right\},\qquad
k=2,3,\qquad N=2,\dots ,\infty. \nonumber\\
%\label{47}
\frac{\partial D^{(2)}(U(x))}{\partial t_{N}}=
-2(2N-1)\frac{\partial D^{(2)N}(U(x))}{\partial x},\\
%\label{48}
\frac{\partial D^{(3)}(U(x))}{\partial t_{N}}=
-6ND^{(3)}(U(x))\frac{\partial D^{(2)N-1}(U(x))}{\partial x}
-2ND^{(2)N-1}(U(x))\frac{\partial D^{(3)}(U(x))}{\partial x}.
\end{gather*}

 The construction of integrable equations with $SU(n)$ symmetries for $n \ge 4$
has dif\/f\/iculties in reduction of non-primitive invariant
currents to primitive currents.

The similar method of construction of chiral currents for
$SO(2l+1)=B_{l}$, $SP(2l)=C_{l}$ groups was used by Evans et
al.~\cite{ger:Eva} on the base of symmetric invariant tensors of
de Azcarraga et al.~\cite{ger:Az}. In the def\/ining
representation these group generators corresponding to algebras
$t_{\mu}$ satisfy the rules
\[ [t_{\mu},t_{\nu}]=2if_{\mu\nu}^{\lambda}t_{\lambda},\qquad
{\rm Tr}(t_{\mu}t_{\nu})=2\delta_{\mu\nu},\qquad t_{\mu}\eta=-\eta
t_{\mu}^{t},\] where $\eta$ is a Euclidean or symplectic
structure.

The symmetric tensor structure constants for these groups were
introduced through completely symmetrized product of three
generators of corresponding algebras
\begin{gather*}%\label{eq49}
t_{(\mu}t_{\nu}t_{\lambda )}= v^{\rho}_{\mu\nu\lambda}t_{\rho},
\end{gather*} where
$v_{\mu\nu\lambda\rho}$ is a totally symmetric tensor.
 The basic invariant
symmetric tensors have the form~\cite{ger:Az}
\begin{gather*}%\label{eq50}
V^{(2)}_{\mu\nu}=\delta_{\mu\nu},\,V^{(2N)}_{(\mu_{1}\mu_{2}\dots
\mu_{2N-1}\mu_{2N})}=
v^{\nu_{1}}_{(\mu_{1}\mu_{2}\mu_{3}}v^{\nu_{1}\nu_{2}}_{\mu_{4}\mu_{5}}\cdots
v^{\nu_{2N-3}}_{\mu_{2N-2}\mu_{2N-1}\mu_{2N})}, \qquad N=2,\dots
,\infty .
\end{gather*} The invariant chiral currents
$J^{(2N)}$ (\ref{eq36}) coincide with the basis invariant chiral
currents $V^{(2N)}$
\begin{gather*}%\label{eq51}
J^{(2N)}=2V^{(2N)}_{\mu{1}\dots \mu_{2N}}U^{\mu _{1}}\cdots
U^{\mu_{2N}}.\end{gather*} The commutation relations of invariant
chiral currents are PBs of hydrodynamic type
\begin{gather}
\{ J^{(M)}(x),J^{(N)}(y)\}=-MN J^{(M+N-2)}(x)\frac{\partial
}{\partial x}\delta(x-y)\nonumber\\
\phantom{\{ J^{(M)}(x),J^{(N)}(y)\}=}{}
-\frac{MN(N-1)}{M+N-2}\frac{\partial J^{(M+N-2)}(x)} {\partial
x}\delta(x-y).\label{eq52}
\end{gather} The
commuting charges of these  invariant chiral currents are
dynamical Casimir operators on $SO(2l+1)$, $SP(2l)$. The metric
tensor of Riemmann space of invariant chiral currents is as
follows
\begin{gather}g_{MN}(J(x))=-MN(M+N-2) J^{(M+N-2)}(x).\nonumber\end{gather}
 The commutation relations (\ref{eq52}) coincide with commutation relations, which was obtained
by Evans at~al.~\cite{ger:Eva}.

 We used relations for new symmetric invariant tensors $V^{(2N,1)}_{(\mu_{1}\dots \mu_{2N})}$ (see Appendix~B),
which we obtained during calculation PB (\ref{eq52})
\begin{gather*}
v_{(\mu_{1}\mu_{2}\mu_{3}}^{k}v_{\mu_{4}\mu_{5}\mu_{6}}^{l}v_{\mu_{7}\mu_{8}\mu_{9}}^{n}v_{\mu_{10})}^{kln}=
V^{(10)}_{(\mu_{1}\dots \mu_{10})},\\
 v_{(\mu_{1}\mu_{2}\mu_{3}}^{k}v_{\mu_{4}\mu_{5}\mu_{6}}^{l}v_{\mu_{7}\mu_{8}\mu_{9}}^{n}
v_{\mu_{10}\mu_{11}\mu_{12})}^{m}v^{klnm}=V^{(12)}_{(\mu_{1}\dots \mu_{12})},\\
 v_{(\mu_{1}\mu_{2}\mu_{3}}^{k}v_{\mu_{4}\mu_{5}\mu_{6}}^{l}v_{\mu_{7}\mu_{8}\mu_{9}}^{n}
v_{\mu_{10}\mu_{11}\mu_{12}}^{m}v_{\mu_{13}}^{klp}v_{\mu_{14})}^{nmp}=
V^{(14)}_{(\mu_{1}\dots \mu_{14})}.
\end{gather*}

\pdfbookmark[1]{Appendix A}{appendixA}
\section*{Appendix A}

 The new dependent invariant chiral currents and the new dependent totally
symmetric invariant tensors for $SU(N)$ group can be obtained
under dif\/ferent order of calculation of trace of the pro\-duct
of the generators of $SU(n)$ algebra. Let us mark the matrix
product of two generators~$t_{\mu}$,~$t_{\nu}$ in round brackets
\begin{gather}\label{eq54}
(t_{\mu}t_{\nu})=\frac{2}{n}\delta_{\mu\nu}+(d^{k}_{\mu\nu}+if^{k}_{\mu\nu})
t_{k}.
\end{gather} The expression of invariant chiral currents
$J_{M}(U)$ depends on the position of the matrix product of two
generators in the general list of generators. For example
\begin{gather*}
 J^{(6)}={\rm Tr}[t(tt)(tt)t]=2D^{(6)}+\frac{4}{n}D^{(3)2}+\frac{8}{n}D^{(2)}D^{(4)}+
\frac{8}{n^{2}}D^{(2)3},\\
 J^{(6)}={\rm Tr}[(tt)(tt)(tt)]=2D^{(6,1)}+\frac{12}{n}D^{(2)}D^{(4)}+\frac{8}{n^{2}}D^{(2)3},
\\
%\label{eq55}
J^{(7)}={\rm
Tr}[t(tt)t(tt)t]=2D^{(7)}+\frac{8}{n}D^{(3)}D^{(4)}+\frac{8}{n^{2}}D^{(2)}D^{(5)}+
\frac{24}{n^{2}}D^{(2)2}D^{(3)},
\\
 J^{(7)}={\rm Tr}[(tt)(tt)(tt)t]=2D^{(7,1)}+\frac{4}{n}D^{(3)}D^{(4)}+\frac{12}{n^{2}}D^{(2)}
D^{(5)}+ \frac{24}{n^{2}}D^{(2)2}D^{(3)},
\\
 J^{(8)}={\rm Tr}[t(tt)tt(tt)t]=2D^{(8)}+\frac{4}{n}D^{(4)2}+\frac{8}{n}D^{(3)}D^{(5)}+
\frac{8}{n}D^{(2)}D^{(6)}\\
\phantom{J^{(8)}=}{}
+\frac{24}{n^{2}}D^{(2)}D^{(3)2}+\frac{24}{n^{2}}D^{(2)2}D^{(4)}+\frac{16}
{n^{3}}D^{(2)4},\\
 J^{(8)}={\rm Tr} [(tt)(tt)t(tt)t]=2D^{(8,1)}+\frac{4}{n}D^{(4)2}+\frac{4}{n}D^{(3)}D^{(5)}+\frac{24}{n^{2}}D^{(2)}
D^{(3)2}\\
\phantom{J^{(8)}=}{} +\frac{12}{n}D^{(2)}D^{(6)}+
\frac{24}{n^{2}}D^{(2)2}D^{(4)}+\frac{16}{n^{3}}D^{(2)4},\\
J^{(8)}={\rm Tr}
[(tt)(tt)(tt)(tt)]=2D^{(8,2)}+\frac{4}{n}D^{(4)2}+\frac{16}{n}D^{(2)}
D^{(6,1)}+\frac{32}{n^{2}}D^{(2)2}D^{(4)}+\frac{16}{n^{3}}D^{(2)4},\\
 J^{(8)}={\rm Tr}[t(tt)(tt)(tt)t]=2D^{(8,3)}+
\frac{12}{n}D^{(2)}D^{(6)}+\frac{8}{n}D^{(3)}D^{(5)}\\
\phantom{J^{(8)}=}{}
+\frac{24}{n^{2}}D^{(2)2}D^{(4)}+\frac{24}{n^{2}}D^{(2)}D^{(3)2}+
\frac{16}{n}D^{(2)4},\\
 J^{(9)}={\rm Tr}[t(tt)ttt(tt)t]=2D^{(9)}+\frac{8}{n}D^{(4)}D^{(5)}+\frac{8}{n}D^{(3)}D^{(6)}+
\frac{8}{n}D^{(2)}D^{(7)}+\frac{8}{n^{2}}D^{(3)3}
\\
\phantom{J^{(9)}=}{}
+\frac{48}{n^{2}}D^{(2)}D^{(3)}D^{(4)}+\frac{24}{n^{2}}D^{(2)2}D^{(5)}+\frac{64}{n^{3}}
D^{(2)3}D^{(3)},\\
 J^{(9)}={\rm Tr}[t(tt)tt(tt)(tt)]\\
 \phantom{J^{(9)}}{} =
 \left\{\!\!\!\begin{array}{l}
 \displaystyle 2D^{(9,1)}+\frac{4}{n}D^{(4)}D^{(5)}+\frac{4}{n}D^{(2)}D^{(7)}+
\frac{4}{n}D^{(2)}D^{(7,1)}+ \frac{8}{n}D^{(3)}D^{(6,1)}\vspace{2mm}\\
\displaystyle
\quad{}+\frac{32}{n^{2}}D^{(2)}D^{(3)}D^{(4)}+\frac{32}{n^{2}}D^{(2)2}D^{(5)}+\frac{64}{n^{3}}
D^{(2)3}D^{(3)}, \vspace{1mm} \\ \cdots\cdots\cdots
\cdots\cdots\cdots \cdots\cdots\cdots \cdots\cdots\cdots
\cdots\cdots\cdots \cdots\cdots\cdots
\vspace{1mm}\\
\displaystyle 2D^{(9,4)}+\frac{4}{n}D^{(2)}D^{(7)}+
\frac{4}{n}D^{(2)}D^{(7,1)}+
\frac{12}{n}D^{(3)}D^{(6,1)}\vspace{2mm}\\
\displaystyle\quad{}
+\frac{32}{n^{2}}D^{(2)}D^{(3)}D^{(4)}+\frac{32}{n^{2}}D^{(2)2}D^{(5)}+\frac{64}{n^{3}}
D^{(2)3}D^{(3)}, \end{array}\right.
\\
J^{(9)}={\rm Tr}[t(tt)t(tt)t(tt)]\\
\phantom{J^{(9)}}{} = \left\{\!\!\!\begin{array}{l} \displaystyle
2D^{(9,2)}+\frac{4}{n}D^{(4)}D^{(5)}+\frac{8}{n}D^{(3)}D^{(6)}+
\frac{8}{n}D^{(2)}D^{(7)}+\frac{4}{n}D^{(2)}D^{(7,1)}\vspace{2mm}\\
\displaystyle\quad{}
+\frac{8}{n^{2}}D^{(3)3}+\frac{40}{n^{2}}D^{(2)}D^{(3)}D^{(4)}+\frac{32}{n^{2}}
D^{(2)2}D^{(5)}+\frac{64}{n^{3}}D^{(2)3}D^{(3)},\vspace{1mm} \\
\cdots\cdots\cdots \cdots\cdots\cdots \cdots\cdots\cdots
\cdots\cdots\cdots \cdots\cdots\cdots \cdots\cdots\cdots
\vspace{1mm}\\
\displaystyle \displaystyle
2D^{(9,3)}+\frac{8}{n}D^{(2)}D^{(7)}+\frac{4}{n}D^{(2)}D^{(71)}+\frac{12}{n}D^{(3)}D^{(6)}+
\frac{8}{n}D^{(3)3}\vspace{2mm}\\
\displaystyle\quad{}+\frac{40}{n^{2}}D^{(2)}D^{(3)}D^{(4)}+\frac{32}{n^{2}}D^{(2)2}D^{(5)}+\frac{64}{n^{3}}
D^{(2)3}D^{(3)}, \end{array}\right.
\end{gather*}
where $t=t_{\mu}U^{\mu}$ and two variants of two last expressions
for $J^{(9)}(U)$ were obtained from two variants of expression for
$J^{(6)}(U)$ during calculation $J^{(9)}(U)$.
 Because the result of calculation does not depend
on the order of calculation, we can obtain relations between new
invariant chiral currents and basic invariant currents
$D^{(M)}(U)$
\begin{gather*}
 D^{(6,1)}=D^{(6)}+\frac{2}{n}D^{(3)2}-\frac{2}{n}D^{(2)}D^{(4)},\\ D^{(7,1)}=D^{(7)}+\frac{4}{n}D^{(3)}D^{(4)}-\frac{4}{n}D^{(2)}D^{(5)},\\
 %\label{eq56}
 D^{(81)}=D^{(8)}+\frac{2}{n}D^{(3)}D^{(5)}-\frac{2}{n}D^{(2)}D^{(6)},\\ D^{(8,2)}=D^{(8)}+\frac{4}{n}D^{(3)}D^{(5)}-\frac{4}{n}D^{(2)}D^{(6)}
 -\frac{4}{n^{2}}D^{(2)}D^{(3)2}+\frac{4}{n^{2}}D^{(2)2}D^{(4)},\\ D^{(8,3)}=D^{(8)}+\frac{2}{n}D^{(4)2}-\frac{2}{n}D^{(2)}D^{(6)},\\ D^{(9,1)}=D^{(9)}+\frac{2}{n}D^{(4)}D^{(5)}-\frac{4}{n^{2}}D^{(3)3}+\frac{8}{n^{2}}D^{(2)}D^{(3)}D^{(4)}+
\frac{4}{n^{2}}D^{(2)2}D^{(5)},\\
D^{(9,2)}=D^{(9)}+\frac{2}{n}D^{(4)}D^{(5)}-\frac{2}{n}D^{(2)}D^{(7)}-\frac{4}{n^{2}}D^{(2)}D^{(3)}D^{(4)}+
\frac{4}{n^{2}}D^{(2)2}D^{(5)},\\
 D^{(9,3)}=D^{(9)}+\frac{4}{n}D^{(4)}D^{(5)}-\frac{2}{n}D^{(2)}D^{(7)}- \frac{2}{n}D^{(3)}D^{(6)}-\frac{4}{n^{2}}D^{(2)}D^{(3)}D^{(4)}+
\frac{4}{n^{2}}D^{(2)2}D^{(5)},\\
D^{(9,4)}=D^{(9)}+\frac{4}{n}D^{(4)}D^{(5)}-\frac{2}{n}D^{(3)}D^{(6)}-\frac{8}{n^{2}}D^{(3)3}
+\frac{12}{n^{2}}D^{(2)}D^{(3)}D^{(4)}+
\frac{4}{n^{2}}D^{(2)2}D^{(5)}.
\end{gather*}

Hence we can obtain the new relations for symmetric tensors
\begin{gather*}
d^{k}_{(\mu\nu}d^{l}_{\lambda\rho}d^{n}_{\sigma\varphi)}d^{kln}
=d^{k}_{(\mu\nu}d^{kl}_{\lambda}d^{ln}_{\rho}d^{n}_{\sigma\varphi)}+
\frac{2}{n}d_{(\mu\nu\lambda}d_{\rho\sigma\varphi)}
-\frac{2}{n}\delta_{(\mu\nu}d^{k}_{\lambda\rho}d^{k}_{\sigma\varphi)},\\
%\label{eq57}
d^{k}_{(\mu\nu}d^{l}_{\lambda\rho}d^{n}_{\sigma\varphi}d^{nm}_{\tau)}d^{klm}=
d^{k}_{(\mu\nu}d^{kl}_{\lambda}d^{ln}_{\rho}d^{nm}_{\sigma}d^{m}_{\varphi\tau)}+
\frac{4}{n}d_{(\mu\nu\lambda}d^{k}_{\rho\sigma}d^{k}_{\varphi\tau)}
-\frac{4}{n}\delta_{(\mu\nu}d^{k}_{\lambda\rho}d^{kl}_{\sigma}
d^{l}_{\varphi\tau)},\\
d^{k}_{(\mu\nu}d^{l}_{\lambda\rho}d^{n}_{\sigma\varphi}d^{nm}_{\tau}d^{mp}_{\tau)}d^{klp}=
d^{k}_{(\mu\nu}d^{kl}_{\lambda}d^{ln}_{\rho}d^{nm}_{\sigma}d^{mp}_{\varphi}d^{p}_{\tau\theta
)}+
\frac{4}{n}d_{(\mu\nu\lambda}d^{k}_{\rho\sigma}d^{kl}_{\varphi}d^{l}_{\tau\theta
)}-
\frac{2}{n}\delta_{(\mu\nu}d^{k}_{\lambda\rho}d^{kl}_{\sigma}d^{ln}_{\varphi}d^{n}_{\tau\theta )},\\
d^{k}_{(\mu\nu}d^{l}_{\lambda\rho}d^{n}_{\sigma\varphi}d^{m}_{\tau\theta)}d^{klp}d^{nmp}=
d^{k}_{(\mu\nu}d^{kl}_{\lambda}d^{ln}_{\rho}d^{nm}_{\sigma}d^{mp}_{\varphi}d^{p}_{\tau\theta)}+
\frac{4}{n}d_{(\mu\nu\lambda}d^{k}_{\rho\sigma}d^{kl}_{\varphi}d^{l}_{\tau\theta)}\\
\phantom{d^{k}_{(\mu\nu}d^{l}_{\lambda\rho}d^{n}_{\sigma\varphi}d^{m}_{\tau\theta)}d^{klp}d^{nmp}=}{}
-\frac{4}{n}\delta_{(\mu\nu}d^{k}_{\lambda\rho}d^{kl}_{\sigma}d^{ln}_{\varphi}
d^{n}_{\tau\theta)}-\frac{4}{n^{2}}\delta_{(\mu\nu}d_{\lambda\rho\sigma}d_{\varphi\tau\theta)}+
\frac{4}{n^{2}}\delta_{(\mu\nu}\delta_{\lambda\rho}d^{k}_{\sigma\varphi}d^{k}_{\tau\theta)}.
\end{gather*}

It is possible to obtain similar relations for invariant symmetric
tensors of ninth order. The commutation relations of chiral
currents in terms of the basic invariant currents are as follows
\begin{gather*}
\{D^{(5)}(x),D^{(3)}(y)\}=-\left[15D^{(6)}(x)+\frac{6}{n}D^{(3)2}(x)-\frac{6}{n}D^{(2)}(x)
D^{(4)}(x)\right]\frac{\partial }{\partial x}\delta(x-y)\\
\phantom{\{D^{(5)}(x),D^{(3)}(y)\}=}{}-\frac{1}{3}\frac{\partial}{\partial
x}\left[15D^{(6)}(x)+\frac{6}{n}D^{(3)2}(x)-\frac{6}{n}
D^{(2)}(x)D^{(4)}(x)\right]\delta(x-y),\\
\{D^{(5)}(x),D^{(4)}(y)\}=-\left[20D^{(7)}(x)+\frac{16}{n}D^{(3)}(x)D^{(4)}(x)-\frac{16}{n}
D^{(2)}(x)D^{(5)}(x)\right]\frac{\partial }{\partial x}\delta(x-y)\\
%\label{eq58}
\phantom{\{D^{(5)}(x),D^{(4)}(y)\}=}{} -\frac{3}{7}\frac{\partial
}{\partial x}
\!\left[20D^{(7)}(x)\!+\frac{16}{n}D^{(3)}(x)D^{(4)}(x)\!-
\frac{16}{n}D^{(2)}(x)D^{(5)}(x)\right]\!\delta(x-y),\\
\{D^{(5)}(x),D^{(5)}(y)\}=-\left[25D^{(8)}(x)+\frac{36}{n}D^{(3)}(x)D^{(5)}(x)-\frac{20}{n}
D^{(2)}(x)D^{(6)}(x)\right.\\
\left. \phantom{\{D^{(5)}(x),D^{(5)}(y)\}=}{}
-\frac{4}{n}D^{(2)}(x)D^{(3)2}(x)
+\frac{4}{n^{2}}D^{(2)2}(x)D^{(4)}(x)\right]\frac{\partial }{\partial x}\delta(x-y)\\
\phantom{\{D^{(5)}(x),D^{(5)}(y)\}=}{}- \frac{1}{2}\frac{\partial
}{\partial x}\left[25D^{(8)}(x) +\frac{36}{n}D^{(3)}(x)D^{(5)}(x)-
\frac{20}{n}D^{(2)}(x)D^{(6)}(x)\right.\\
\left.\phantom{\{D^{(5)}(x),D^{(5)}(y)\}=}{}
-\frac{4}{n}D^{(2)}(x)D^{(3)2}(x)+\frac{4}{n^{2}}D^{(2)2}(x)D^{(4)}(x)\right],\\
\{D^{(6)}(x),D^{(4)}(y)\}=-\left[24D^{(8)}+\frac{12}{n}D^{(4)2}-\frac{12}{n}D^{(2)}D^{(6)}\right]
\frac{\partial }{\partial x}\delta(x-y)\\
\phantom{\{D^{(6)}(x),D^{(4)}(y)\}=}{} - \frac{3}{8}\frac{\partial
}{\partial x}
\left[24D^{(8)}+\frac{12}{n}D^{(4)2}-\frac{12}{n}D^{(2)}D^{(6)}\right]\delta(x-y),\\
\{D^{(7)}(x),D^{(3)}(y)\}=-\left[21D^{(8)}+\frac{6}{n}D^{(4)2}+\frac{12}{n}D^{(3)}D^{(5)}
-\frac{18}{n}D^{(2)}D^{(6)}\right] \frac{\partial }{\partial x}\delta(x-y)\\
\phantom{\{D^{(7)}(x),D^{(3)}(y)\}=}{} -\frac{1}{4}\frac{\partial
}{\partial x}
\left[21D^{(8)}+\frac{6}{n}D^{(4)2}+\frac{12}{n}D^{(3)}D^{(5)}-\frac{18}{n}D^{(2)}D^{(6)}\right]\delta(x-y),\\
\{D^{(8)}(x),D^{(3)}(y)\}=-\left[24D^{(9)}-\frac{12}{n}D^{(2)}D^{(7)}+\frac{24}{n}D^{(4)}D^{(5)}-\frac{24}{n^{2}}D^{(3)3}+
\frac{24}{n^{2}}D^{(2)}D^{(3)}D^{(4)}\!\!\right.\\
\left. \phantom{\{D^{(8)}(x),D^{(3)}(y)\}=}{} +
\frac{48}{n^{2}}D^{(2)2}D^{(5)}\right]\frac{\partial }{\partial
x}\delta(x-y)
-\frac{2}{9}\frac{\partial }{\partial x}\left[24D^{(9)}-\frac{12}{n}D^{(2)}D^{(7)}\right.\\
\left.\phantom{\{D^{(8)}(x),D^{(3)}(y)\}=}{}
+\frac{24}{n}D^{(4)}D^{(5)}\!- \frac{24}{n^{2}}D^{(3)3}\!+
\frac{24}{n^{2}}D^{(2)}D^{(3)}D^{(4)}\!+\frac{48}{n^{2}}D^{(2)2}D^{(5)}\right]\!\delta(x-y),\\
\{D^{(7)}(x),D^{(4)}(y)\}=-\left[28D^{(9)}-\frac{8}{n}D^{(3)}D^{(6)}-\frac{24}{n}D^{(2)}D^{(7)}+\frac{32}{n}D^{(4)}D^{(5)}
\right.\\
\left. \phantom{\{D^{(7)}(x),D^{(4)}(y)\}=}{} -\frac{48}{n^{2}}D^{(2)}D^{(3)}D^{(4)}+ \frac{48}{n^{2}}D^{(2)2}D^{(5)}\right]\frac{\partial }{\partial x}\delta(x-y)\\
\phantom{\{D^{(7)}(x),D^{(4)}(y)\}=}{} -\frac{1}{3}\frac{\partial
}{\partial x}\left[28D^{(9)}-\frac{8}{n}D^{(3)}D^{(6)}
- \frac{24}{n}D^{(2)}D^{(7)}+\frac{32}{n}D^{(4)}D^{(5)}\right.\\
\left. \phantom{\{D^{(7)}(x),D^{(4)}(y)\}=}{}-
\frac{48}{n^{2}}D^{(2)}D^{(3)}D^{(4)}+\frac{48}{n^{2}}D^{(2)2}D^{(5)}\right]\delta(x-y),\\
\{D^{(6)}(x),D^{(5)}(y)\}=-\left[30D^{(9)}-\frac{4}{n}D^{(3)}D^{(6)}-\frac{12}{n}D^{(2)}D^{(7)}+\frac{32}{n}D^{(4)}D^{(5)}-
\frac{32}{n^{2}}D^{(3)3}\right.\\
\left. \phantom{\{D^{(6)}(x),D^{(5)}(y)\}=}{} + \frac{24}{n^{2}}D^{(2)}D^{(3)}D^{(4)}+\frac{56}{n^{2}}D^{(2)2}D^{(5)}\right]\frac{\partial }{\partial x}\delta(x-y)\\
\phantom{\{D^{(6)}(x),D^{(5)}(y)\}=}{}-\frac{4}{9}\frac{\partial
}{\partial x}\left[30D^{(9)}
 -\frac{4}{n}D^{(3)}D^{(6)}
-\frac{12}{n}D^{(2)}D^{(7)}+\frac{32}{n}D^{(4)}D^{(5)}\right.\\
\left.\phantom{\{D^{(6)}(x),D^{(5)}(y)\}=}{} -
\frac{32}{n^{2}}D^{(3)3}+\frac{24}{n^{2}}D^{(2)}D^{(3)}D^{(4)}+\frac{56}{n^{2}}D^{(2)2}D^{(5)}\right]\delta(x-y).
\end{gather*}

\pdfbookmark[1]{Appendix B}{appendixB}
\section*{Appendix B}

 The invariant chiral currents $J^{(2N)}$ and $V^{(2N)}$ and the new dependent totally
symmetric invariant tensors for $SO(2l+1)$, $SP(2l)$ groups can be
obtained under dif\/ferent order of calculation of trace of the
product of the generators of corresponding algebras.
 Let us mark the matrix product of three generators $t_{\mu}$ in round brackets
\[(t_{(\mu}t_{\nu}t_{\lambda)})=v_{\mu\nu\lambda\rho}t_{\rho}.\]
A dif\/ferent position of this triplet inside of $J^{2N}$ produces
dif\/ferent expressions for $V^{2N}$
\begin{gather*}
J^{(10)}={\rm
Tr}[((t_{1}t_{2}t_{3})t_{4}(t_{5}t_{6}t_{7})(t_{8}t_{9}t_{10}))]U_{1}\cdots
U_{10}=
2v^{k}_{123}v^{kl}_{45}v^{ln}_{67}v^{n}_{8910}U_{1}\cdots U_{10}=2V^{(10)},\\
J^{(10)}={\rm
Tr}[((t_{1}t_{2}t_{3})(t_{4}t_{5}t_{6})(t_{7}t_{8}t_{9})t_{10})]U_{1}\cdots
U_{10}=
2v^{k}_{123}v^{l}_{456}v^{n}_{789}v^{kln}_{10}U_{1}\cdots U_{10}=2V^{(10,1)},\\
J^{(12)}={\rm Tr} [(t_{1}(t_{2}t_{3}t_{4})t_{5}(t_{6}t_{7}t_{8})t_{9}(t_{10}t_{11}t_{12}))]U_{1}\cdots U_{12}\\
\phantom{J^{(12)}}{}=
2v^{k}_{123}v^{kl}_{45}v^{ln}_{67}v^{nm}_{89}v^{m}_{101112}U_{1}\cdots U_{12}=2V^{(12)},\\
J^{(12)}={\rm Tr}[((t_{1}t_{2}t_{3})(t_{4}t_{5}t_{6})(t_{7}t_{8}t_{9})(t_{10}t_{11}t_{12}))]U_{1}\cdots U_{12}\\
\phantom{J^{(12)}}{} =2v^{k}_{123}v^{l}_{456}v^{n}_{789}v^{m}_{101112}v^{klnm}U_{1}\cdots U_{12}=2V^{(12,1)},\\
J^{(14)}={\rm Tr}[((t_{1}t_{2}t_{3})t_{4}(t_{5}t_{6}t_{7})t_{8}(t_{9}t_{10}t_{11})(t_{12}t_{13}t_{14}))]U_{1}\cdots U_{14}\\
\phantom{J^{(14)}}{} =2v^{k}_{123}v^{kl}_{45}v^{ln}_{67}v^{nm}_{89}v^{mp}_{1011}v^{p}_{121314}U_{1}\cdots U_{14}=2V^{(14)},\\
J^{(14)}={\rm Tr} [((t_{1}t_{2}t_{3})(t_{4}t_{5}t_{6})(t_{7}t_{8}t_{9})(t_{10}t_{11}t_{12})t_{13}t_{14})]U_{1}\cdots U_{14}\\
\phantom{J^{(14)}}{} =
2v^{k}_{123}v^{l}_{456}v^{n}_{789}v^{m}_{101112}v^{klp}_{13}v^{nmp}_{14}U_{1}\cdots
U_{14}=2V^{(14,1)}.
\end{gather*}
 Here we introduced the short notation $t_{\mu_{k}}=t_{k}$, $U^{\mu_{k}}=U_{k}$ and $v^{k}_{\mu_{l}\mu_{n}\mu_{m}}$=$v^{k}_{lnm}$.
New invariant chiral tensors do not lead to new invariant chiral
currents.

\subsection*{Acknowledgments}

The author would like to thank J.A. de Azcarraga for the
stimulating discussion about non-primitive invariant tensors on
simple Lie algebras.

\pdfbookmark[1]{References}{ref}
\LastPageEnding

\end{document}